\documentclass[%
reprint,
superscriptaddress,
showpacs,
amsmath,amssymb,
prc,
floatfix, ]%
{revtex4-1}

\usepackage{color}

\usepackage{graphicx}% Include figure files
\usepackage{dcolumn}% Align table columns on decimal point
\usepackage{bm}% bold math
\usepackage[dvipdfmx,bookmarks=true,colorlinks,%
            citecolor=blue,linkcolor=blue,anchorcolor=blue,filecolor=blue,urlcolor=blue,%
           ]{hyperref}
\allowdisplaybreaks

\begin{document}

%\begin{CJK*}{GBK}{}

\title{Multidimensionally-constrained relativistic mean-field study of triple-humped barriers in actinides}

\author{Jie Zhao}% (ÕÔ½Ü)}%
%\email{zhaojie@itp.ac.cn}
 \affiliation{State Key Laboratory of Theoretical Physics,
              Institute of Theoretical Physics, Chinese Academy of Sciences, Beijing 100190, China}
\author{Bing-Nan Lu}% (ÂÀ±þéª)}%
%\email{bnlu@itp.ac.cn}
 \affiliation{State Key Laboratory of Theoretical Physics,
              Institute of Theoretical Physics, Chinese Academy of Sciences, Beijing 100190, China}
 \affiliation{Institut f\"ur Kernphysik (IKP-3) and J\"ulich Center for Hadron Physics,
              Forschungszentrum J\"ulich, D-52425 J\"ulich, Germany}
\author{Dario Vretenar}%
%\email{vretenar@phy.hr}
 \affiliation{Physics Department, Faculty of Science, University of Zagreb, Bijenicka 32,
              Zagreb 10000, Croatia}
\author{En-Guang Zhao}% (ÕÔ¶÷¹ã)}%
%\email{egzhao@mail.itp.ac.cn}
 \affiliation{State Key Laboratory of Theoretical Physics,
              Institute of Theoretical Physics, Chinese Academy of Sciences, Beijing 100190, China}
 \affiliation{Center of Theoretical Nuclear Physics, National Laboratory
              of Heavy Ion Accelerator, Lanzhou 730000, China}
\author{Shan-Gui Zhou}% (ÖÜÉƹó)}%
 \email{sgzhou@itp.ac.cn}
%\homepage{http://www.itp.ac.cn/~sgzhou}
 \affiliation{State Key Laboratory of Theoretical Physics,
              Institute of Theoretical Physics, Chinese Academy of Sciences, Beijing 100190, China}
 \affiliation{Center of Theoretical Nuclear Physics, National Laboratory
              of Heavy Ion Accelerator, Lanzhou 730000, China}
 \affiliation{Center for Nuclear Matter Science, Central China Normal University, Wuhan 430079, China}

\date{\today}

\begin{abstract}
\begin{description}
\item[Background]
Potential energy surfaces (PES's) of actinide nuclei are characterized
by a two-humped barrier structure.
At large deformations beyond the second barrier the occurrence of a third one 
was predicted by macroscopic-microscopic model calculations in the 1970s, 
but contradictory results were later reported by number of studies 
that used different methods.
\item[Purpose]
Triple-humped barriers in actinide nuclei are investigated in the framework 
of covariant density functional theory (CDFT).
\item[Methods]
Calculations are performed using the
multidimensionally-constrained relativistic mean field (MDC-RMF) model,
with the nonlinear point-coupling functional PC-PK1
and the density-dependent meson exchange functional DD-ME2
in the particle-hole channel.
Pairing correlations are treated in the BCS approximation with
a separable pairing force of finite range.
\item[Results]
Two-dimensional PES's of
$^{226,228,230,232}$Th and $^{232,234,236,238}$U are mapped
and the third minima on these surfaces are located. 
Then one-dimensional potential energy curves along 
the fission path are analyzed in detail and the energies of the second barrier, 
the third minimum, and the third barrier are determined. 
The functional DD-ME2 predicts the occurrence of a third barrier 
in all Th nuclei and $^{238}$U.
The third minima in $^{230,232}$Th are very shallow, 
whereas those in $^{226,228}$Th and $^{238}$U are quite prominent.
With the functional PC-PK1 a third barrier is found only in $^{226,228,230}$Th.
Single-nucleon levels around the Fermi surface are analyzed in $^{226}$Th,
and it is found that the formation of the third minimum is mainly due to
the $Z=90$ proton energy gap at $\beta_{20} \approx 1.5$ and $\beta_{30} \approx 0.7$.
\item[Conclusions]
The possible occurrence of a third barrier on the PES's of actinide nuclei 
depends on the effective interaction used in multidimensional CDFT calculations.
More pronounced minima are predicted by the DD-ME2 functional, 
as compared to the functional PC-PK1.
The depth of the third well in Th isotopes decreases with increasing neutron number.
%The origin of the third minimum can be traced to large
%correlations between several pairs of single-proton states around the Fermi surface.
The origin of the third minimum is due to the proton $Z=90$ shell gap 
at relevant deformations.
\end{description}
\end{abstract}

%21.60.Jz       Nuclear Density Functional Theory and extensions
%               (includes Hartree-Fock and random-phase approximations)
%24.75.+i       General properties of fission
%25.85.-w       Fission reactions
%27.90.+b       A >= 220

\maketitle

%\end{CJK*}

\section{Introduction~\label{sec:Introduction}}

The potential energy surface (PES) of a fissile nucleus
plays a crucial role in fission studies.
Observables such as the fragment mass distribution and the fission half-life
are closely related to properties of the PES of a compound nucleus.
The PES's of actinide nuclei are characterized by a two-humped barrier structure
along the static fission path,
which has extensively been studied both experimentally and using a variety of 
theoretical models. We refer the reader, for instance, to
Refs.~\cite{Lu2014_PRC89-014323,Xie2014_SciChinaPMA57-189,
Zhang2014_PRC90-054313,Giuliani2014_PRC90-054311}
and references therein.
Macroscopic-microscopic model calculations predicted, already in the 1970s 
\cite{Moeller1972_NPA192-529, Moeller1973_IAEA-SM-174-202, Moeller1974_NPA229-269}, 
the occurrence of shallow third minima beyond the second barrier.
These minima were used to explain the thorium anomaly \cite{Moeller1973_IAEA-SM-174-202,
Moeller1974_NPA229-269,Bhandari1989_PRC39-917}.
High resolution fission cross section measurements for $^{230-233}$Th
and $^{237}$U supported the existence of shallow third minima
on the PES's of these nuclei \cite{Blons1975_PRL35-1749,Blons1978_PRL41-1282,
Blons1988_NPA477-231,Knowles1982_PLB116-315,Findlay1986_NPA458-217,
Zhang1986_PRC34-1397,Yoneama1996_NPA604-263}.
The deduced values for the depth of the third well are only a few hundred keV.
In Ref.~\cite{Sin2006_PRC74-014608} a model was developed
to describe fission in light actinides, and to consider
transmission through a triple-humped fission barrier with absorption.
Using this model the complex resonance structure in the first-chance
neutron-induced fission cross sections of $^{232}$Th and $^{231}$Pa
was reproduced, and shallow third minima with a depth of less than 1 MeV
were obtained.

The PES's of nuclei in the Ra-Th region were computed using 
the macroscopic-microscopic model with a modified oscillator potential, 
and in many nuclei a third minimum was found at very large quadrupole 
deformation \cite{Bengtsson1987_NPA473-77}.
It was concluded that the depth of the third minimum could be at most 1.5 MeV.
This model, with shell corrections calculated by adopting an axially-deformed
Woods-Saxon potential, was later used to systematically study the PES's of 
even-even Rn, Ra, Th, and U isotopes \cite{Cwiok1994_PLB322-304}.
Very deep minima, or even two hyperdeformed minima, were predicted in many of these nuclei.
In some cases the depth of the third minimum could be as large as 4 MeV.

A series of experiments were performed to find evidence for hyperdeformed states 
in U isotopes \cite{Krasznahorkay1998_PRL80-2073,Krasznahorkay1999_PLB461-15,
Csatlos2005_PLB615-175,Csige2009_PRC80-011301R,Csige2013_PRC87-044321}.
The deduced values for the excitation energies of the third minima and 
the third barriers were $E_{\rm III}=3$--4 MeV and $B_{\rm III} \approx 6$ MeV, respectively.
Thus, the depth of the third well could be 2--3 MeV.
Such deep third minima in U isotopes agree with the predictions
of Ref.~\cite{Cwiok1994_PLB322-304},
but differ from the experimental results of $^{230-233}$Th and
$^{237}$U \cite{Blons1975_PRL35-1749,Blons1978_PRL41-1282,
Blons1988_NPA477-231,Knowles1982_PLB116-315,Findlay1986_NPA458-217,
Zhang1986_PRC34-1397,Yoneama1996_NPA604-263},
and from the theoretical predictions reported in Refs.~\cite{Moeller1972_NPA192-529,
Moeller1973_IAEA-SM-174-202,Moeller1974_NPA229-269,Bengtsson1987_NPA473-77}.

To verify the predictions for third minima in actinides 
in the macroscopic-microscopic model based on the Woods-Saxon potential, 
calculations with additional shape degrees of freedom were preformed in 
Refs.~\cite{Kowal2012_PRC85-061302R,Jachimowicz2013_PRC87-044308}.
It was found that the third barrier could be lowered substantially by including
the $\beta_1$ deformation. Consequently, third minima disappeared in many nuclei,
except for $^{230,232}$Th, in which only a shallow third minimum with 
a depth of less than 400 keV was found.
Furthermore, in Ref.~\cite{Ichikawa2013_PRC87-054326} an analysis of PES's 
computed with the finite-range liquid-drop model \cite{Moeller2009_PRC79-064304}
revealed that only few nuclei accessible to experiment exhibit third minima 
in their PES's and the depth of the third well is less than 1 MeV for 
the light Th and U isotopes.

In addition to the macroscopic-microscopic model, a number of self-consistent approaches 
have also been used to investigate PES's of deformed nuclei.
Deformation-constrained Hartree-Fock or Hartree-Fock-Bogoliubov calculations 
with Skyrme forces \cite{Bonneau2004_EPJA21-391,Samyn2005_PRC72-044316,McDonnell2013_PRC87-054327}
and the Gogny force \cite{Berger1989_NPA502-85,Delaroche2006_NPA771-103,Dubray2008_PRC77-014310}
did not exhibit deep third minima in actinide nuclei.
In Ref.~\cite{Rutz1995_NPA590-680} a shallow third minimum with a depth of 1--2 MeV
in $^{232}$Th was found in axially-deformed relativistic mean-field calculations
with the effective interactions PL-40, NL1, and NL-SH.

In the present study we examine the occurrence and properties of
third minima on the PES's of light actinides, using the
multidimensionally-constrained relativistic mean-field (MDC-RMF) model.
Third minima, if they exist, are located at very large quadrupole and
octupole deformations. It is probable that some
intruder high-$N$ single-particle states play an important role in the
formation of these minima.
Therefore, in addition to the PES's, we analyze in detail
the single-particle level structure at hyperdeformation,
and study the origin of possible third minima.

The paper is organized as follows.
The MDC-RMF model is introduced in Sec.~\ref{sec:model}.
In Sec.~\ref{sec:results} we present numerical details and the
results for the PES's and third barriers in light actinide nuclei.
A summary and conclusions are given in Sec.~\ref{sec:summary}.

\section{\label{sec:model}The MDC-RMF approach}

Relativistic mean-field (RMF) models have been applied very successfully in
studies of a variety of nuclear structure phenomena. 
In the standard representation of RMF models a nucleus is described as a system of 
Dirac nucleons coupled to exchange mesons through an effective Lagrangian
\cite{Serot1986_ANP16-1,
Reinhard1989_RPP52-439, Ring1996_PPNP37-193, Bender2003_RMP75-121,
Vretenar2005_PR409-101, Meng2006_PPNP57-470, Paar2007_RPP70-691,
Niksic2011_PPNP66-519, Qu2013_SciChinaPMA56-2031}.
At the energy scale characteristic for nuclear binding and low-lying excited
states, the meson exchange is just a convenient representation of 
the effective nuclear interaction, and can be replaced by the corresponding 
local four-point (contact) interactions between nucleons \cite{Nikolaus1992_PRC46-1757,
Burvenich2002_PRC65-044308}.
A quantitative treatment of nuclear matter and finite nuclei necessitates 
a medium dependence of effective mean-field interactions
that takes into account higher-order many-body effects. 
The medium dependence can be introduced either by including nonlinear terms
in the Lagrangian \cite{Boguta1977_NPA292-413, Brockmann1992_PRL68-3408, 
Sugahara1994_NPA579-557} or by assuming an explicit density dependence 
for the meson-nucleon couplings \cite{Fuchs1995_PRC52-3043,
Niksic2002_PRC66-024306}. 
Therefore, the framework of covariant density functionals
can be formulated in one of the following four representations for the effective nuclear
interaction: meson exchange (ME) or point-coupling (PC) nucleon interactions combined with
nonlinear (NL) or density-dependent (DD) vertex functions.

In a self-consistent mean-field approach that allows breaking
both axial quadrupole and reflection symmetries, the MDC-RMF model has recently been
developed and applied to studies of PES's and fission barriers of
actinides \cite{Lu2012_PRC85-011301R,Lu2012_EPJWoC38-05003,Lu2014_PRC89-014323,
Lu2014_JPCS492-012014, Lu2014_PS89-054028}, %Lu2014_PST},
shapes of hypernuclei \cite{Lu2011_PRC84-014328,Lu2014_PRC89-044307},
and nonaxial-octupole $Y_{32}$ correlations in $N = 150$ isotones \cite{Zhao2012_PRC86-057304}.
The nuclear shape is parameterized by the deformation parameters
\begin{equation}
 \beta_{\lambda\mu} = {4\pi \over 3AR^\lambda} \langle Q_{\lambda\mu} \rangle,
 \label{eq:01}
\end{equation}
where $Q_{\lambda\mu} = r^\lambda Y_{\lambda \mu}$ is the mass multipole operator.
The shape is invariant under the exchange of the $x$ and $y$ axes and,
thus, all deformations $\beta_{\lambda\mu}$
with even $\mu$ can be included simultaneously.
The deformed RMF equations are solved by an expansion in the axially-deformed 
harmonic oscillator (ADHO) basis \cite{Gambhir1990_APNY198-132, Ring1997_CPC105-77}. 
For details of the MDC-RMF model, we refer the reader to Ref.~\cite{Lu2014_PRC89-014323}.

\section{\label{sec:results}Results and discussion}

\subsection{\label{subsec:checks}Implementation and numerical details}

In the present study we employ two relativistic density functionals:
PC-PK1 \cite{Zhao2010_PRC82-054319} with point-coupling effective
interactions that include nonlinear terms in the self-energies
and the meson-exchange functional DD-ME2 \cite{Lalazissis2005_PRC71-024312}
with density-dependent meson-nucleon couplings.
Pairing correlations are taken into account in the BCS approximation 
with a separable pairing force of finite range \cite{Tian2009_PLB676-44, 
Tian2009_PRC80-024313, Niksic2010_PRC81-054318}
\begin{equation}
 V( \bm{r}_1 - \bm{r}_2 ) = 
  -{g} \delta( \tilde{\bm{R}} - \tilde{\bm{R}^\prime} )
   P(\tilde{\bm{r}}) P(\tilde{\bm{r}^\prime}) \frac{1-\hat{P}_\sigma}{2} ,
\end{equation}
where $g$ is the pairing strength and $\tilde{\bm{R}}$ and $\tilde{\bm{r}}$ 
are the center-of-mass and relative coordinates of the two nucleons, respectively.
$P(\bm{r})$ denotes the Gaussian function:
\begin{equation}
 P(\bm{r}) = \frac{1}{(4\pi a^2)^{3/2}} e^{-{r^2}/{4 a^2}},
\end{equation}
where $a$ is the effective range of the pairing force.
The two parameters $g=g_0 =728$ MeV$\cdot$fm$^3$ and $a=0.644$ fm 
\cite{Tian2009_PLB676-44, Tian2009_PRC80-024313}
have been adjusted to reproduce the density dependence of the 
pairing gap at the Fermi surface
in symmetric nuclear matter and calculated with the Gogny force D1S. 

\begin{figure}
 \includegraphics[width=0.45\textwidth]{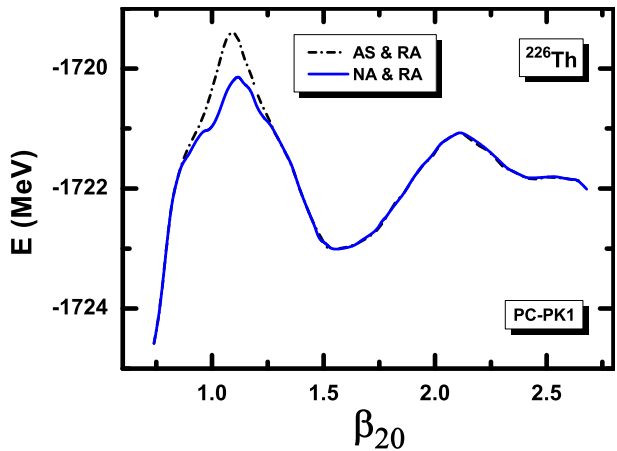}
\caption{(Color online)~\label{fig:Th226Path}%
Energy curve of $^{226}$Th computed with the MDC-RMF model using 
the functional PC-PK1. When axial symmetry is imposed, the energy denoted 
by the dash-dotted (black) curve is obtained  
[axial symmetric (AS) and reflection asymmetric (RA)],
whereas triaxial calculations yield the solid (blue) curve 
[non-axial (NA) and reflection asymmetric (RA)].
The ADHO basis contains $N_{f} = 16$ oscillator shells.
}
\end{figure}

Nonaxial shapes are crucial for determining the height of
both the inner \cite{Abusara2010_PRC82-044303, Abusara2012_PRC85-024314}
and outer barriers \cite{Lu2012_PRC85-011301R, Lu2012_EPJWoC38-05003,
Lu2014_PRC89-014323, Lu2014_JPCS492-012014, Lu2014_PS89-054028}. %Lu2014_PST}.
What role do they play at the third barrier?
In Fig.~\ref{fig:Th226Path} we display the energy curve of $^{226}$Th
computed with and without the inclusion of triaxial deformations. 
This calculation is performed in an ADHO basis truncated to
$N_{f} = 16$ oscillator shells
(see Ref.~\cite{Lu2014_PRC89-014323} for the explanation of the truncation).
As shown in the figure, triaxial deformations lower the second barrier considerably.
Beyond the second saddle point the influence of nonaxial deformations
on the binding energy gradually vanishes as $\beta_{20}$ increases, 
and these shapes appear not to be important at the third minimum and the third barrier.

\begin{figure}
 \includegraphics[width=0.45\textwidth]{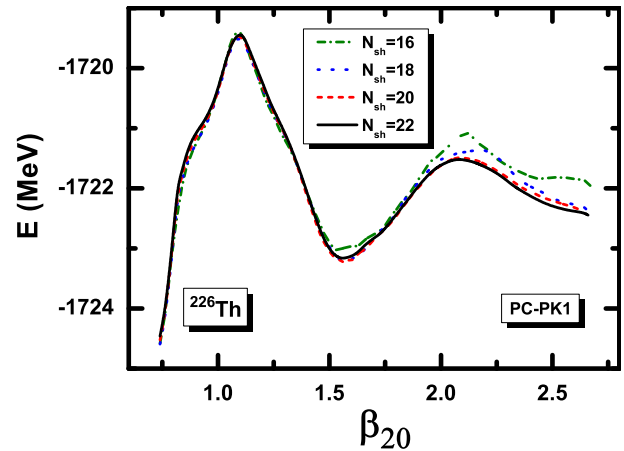}
\caption{(Color online)~\label{fig:Nsh}%
The axially symmetric energy curve of $^{226}$Th computed with the MDC-RMF
model, using the functional PC-PK1. Results obtained with different truncations of the
ADHO basis, i.e., with $N_{f}=16$, 18, 20, and 22 shells are plotted by the dash-dotted,
dotted, dashed, and solid curves, respectively. 
}
\end{figure}

In Ref.~\cite{Lu2014_PRC89-014323} the convergence of the total energy
with respect to the size of the ADHO basis has been verified along the fission path
up to the second fission barrier.
Since here we consider a region with even larger deformation, we will extend 
this check and examine the basis truncation up to the third fission barrier.
The energy curve of $^{226}$Th is calculated up to $\beta_{20}=2.7$ assuming 
axial symmetry and employing ADHO bases with $N_{f}=16$, 18, 20, and 22 shells.
In Ref.~\cite{Lu2014_PRC89-014323} it has been shown that a basis with $N_f=20$
produces very accurate results for the height of the second barrier, around which
the triaxial deformation is also important.
In Fig.~\ref{fig:Th226Path} one notices that nonaxial shapes do not influence the
height of the third barrier, and so the axial symmetry is imposed and 
we mainly focus on the deformation region beyond the second barrier.
Figure~\ref{fig:Nsh} shows that the height of the third barrier
is lowered when $N_f$ increases from 16 to 20.
The results obtained with $N_{f}=20$ and 22 are almost identical, and
this means that $N_{f}=20$ should present an sufficient choice. 
We notice that for the hyperdeformed minimum around $\beta_{20}\sim 1.6$,
results obtained with $N_{f} = 18$, 20, and 22 are difficult to differentiate.

\subsection{Two-dimensional PES's}

\begin{figure*}
 \includegraphics[width=0.8\textwidth]{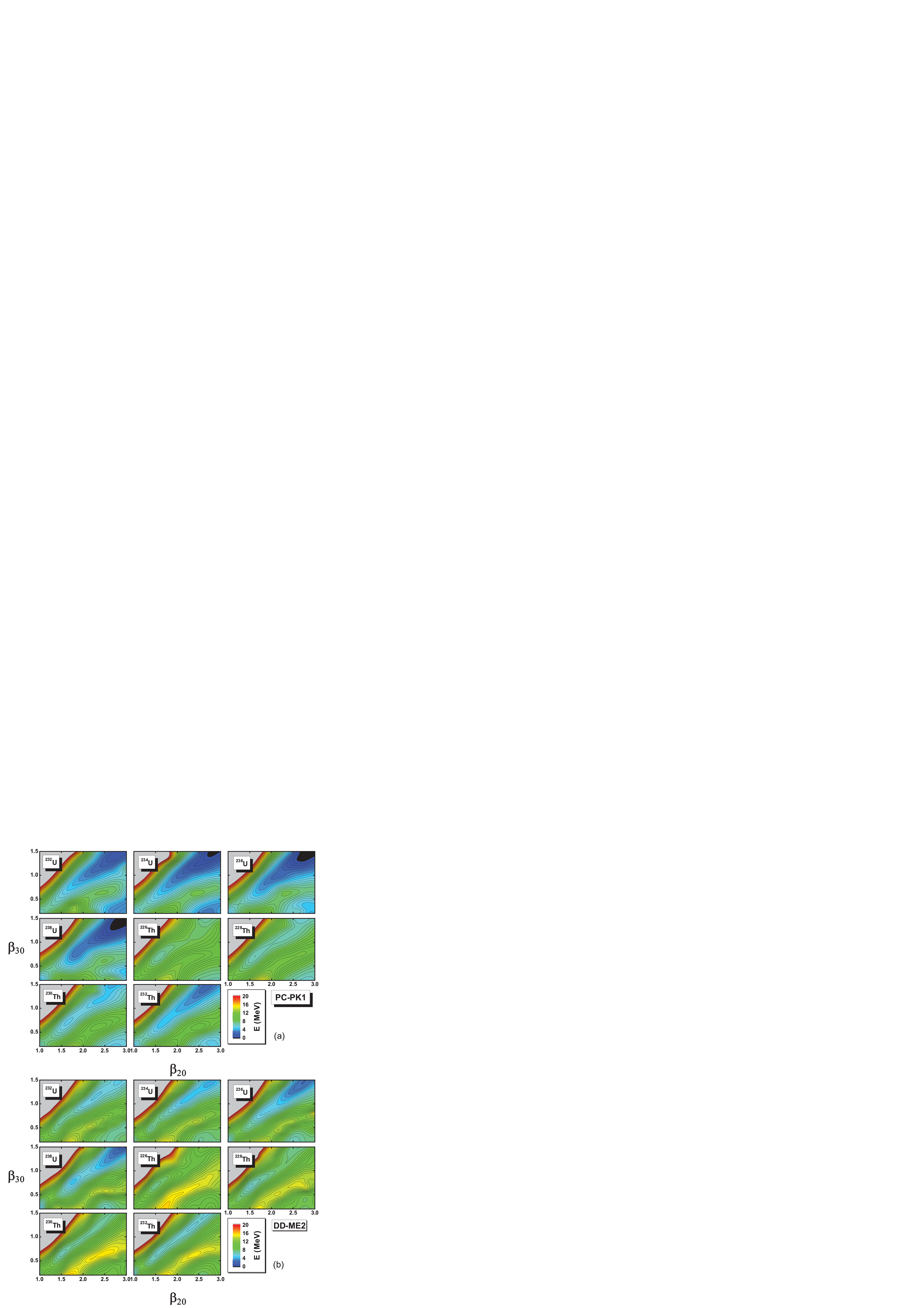}
\caption{(Color online)~\label{fig:pes}%
Self-consistent MDC-RMF energy surfaces of U and Th isotopes in the 
$(\beta_{20},\beta_{30})$ plane, calculated with the relativistic functionals 
PC-PK1 (a) and DD-ME2 (b).
For each nucleus the energy is normalized with respect to the binding energy
of the absolute minimum. The contours join points on the surface with
the same energy and the energy difference between neighboring contours is 0.5 MeV.
The calculation has been performed in the ADHO basis with $N_{f}=16$ shells.
}
\end{figure*}

We consider light even-even actinides
$^{232,234,236,238}$U
and
$^{226,228,230,232}$Th.
Figure~\ref{fig:pes} displays the self-consistent MDC-RMF energy surfaces
in the $(\beta_{20},\beta_{30})$ plane,
calculated with the relativistic functionals PC-PK1 and DD-ME2.
The deformation parameters are restricted to the range $\beta_{20} \sim (1.0, 3.0)$ and
$\beta_{30} \sim (0.3, 1.5)$,
in which the second barrier together with the third minimum and the third barrier,
if they exist, are located.
Note that the contour interval for the PES's plotted in Fig.~\ref{fig:pes} is 0.5 MeV.

In Fig.~\ref{fig:pes}(a) one notices that for $^{232,234,236,238}$U the PES's
do not display a third minimum when calculated with PC-PK1.
For these four U isotopes the second saddle point is located at
$\beta_{20}\sim$ 1.2--1.3 and $\beta_{30}\sim$ 0.3--0.4.
With the increase of $\beta_{20}$ the total binding energy decreases
monotonically along the lowest fission path.
%If exist, the depths of these minima will not exceed 0.5 MeV.
Different results are obtained with DD-ME2, as shown in Fig.~\ref{fig:pes}(b).
With the exception of $^{236}$U, we find a third minimum on the PES's of $^{232,234,238}$U.
The minimum is rather shallow for $^{232}$U and $^{234}$U %, somewhat deeper for $^{234}$U,
and very pronounced for $^{238}$U.
These results are very similar to those obtained in recent calculations based on
the macroscopic-microscopic model \cite{Kowal2012_PRC85-061302R,Jachimowicz2013_PRC87-044308,
Ichikawa2012_PRC86-024610,Ichikawa2013_PRC87-054326}
and the Skyrme Hartree-Fock-Bogoliubov model \cite{McDonnell2013_PRC87-054327}.

In the case of Th isotopes, both PC-PK1 and DD-ME2 predict a pronounced third minimum
for $^{226}$Th with a depth of 2--3 MeV.
As the neutron number $N$ increases, both the energy of the third minimum and the height
of the third barrier decrease and the depth of the third well is also reduced.
This trend has also been predicted in the macroscopic-microscopic model 
calculations \cite{Ichikawa2013_PRC87-054326}.
For $^{230}$Th only a shallow minimum of depth less than 1 MeV occurs.
The third minimum completely disappears for $^{232}$Th when calculated with PC-PK1 
and it is very shallow with the functional DD-ME2.

Figure~\ref{fig:pes} clearly shows that the functional DD-ME2 (lower panel) predicts 
more pronounced third minima when compared to those obtained with PC-PK1 (upper panel). 
In addition to the lowest fission path on which we focus in this work, 
there are other possible paths along which there are more shallow minima and saddle points.
It would be worthwhile to investigate these fine structures in a future study.

\subsection{\label{sec:pairing_strength}Pairing strength}

In general the height of fission barriers is rather sensitive to the strength
of the pairing interaction \cite{Karatzikos2010_PLB689-72}. 
As explained above, the parameters of the separable pairing force of finite range 
that we use in this work were originally adjusted to reproduce the pairing gap 
at the Fermi surface in symmetric nuclear matter and calculated with the Gogny force D1S. 
A number of studies based on the relativistic Hartree-Bogoliubov model have shown 
that this pairing force can be used to calculate structure properties 
with a success; but in some other cases, the pairing strength needs to be fine-tuned 
\cite{Wang2013_PRC87-054331, Afanasjev2013_PRC88-014320}. 

\begin{figure}
 \includegraphics[width=0.45\textwidth]{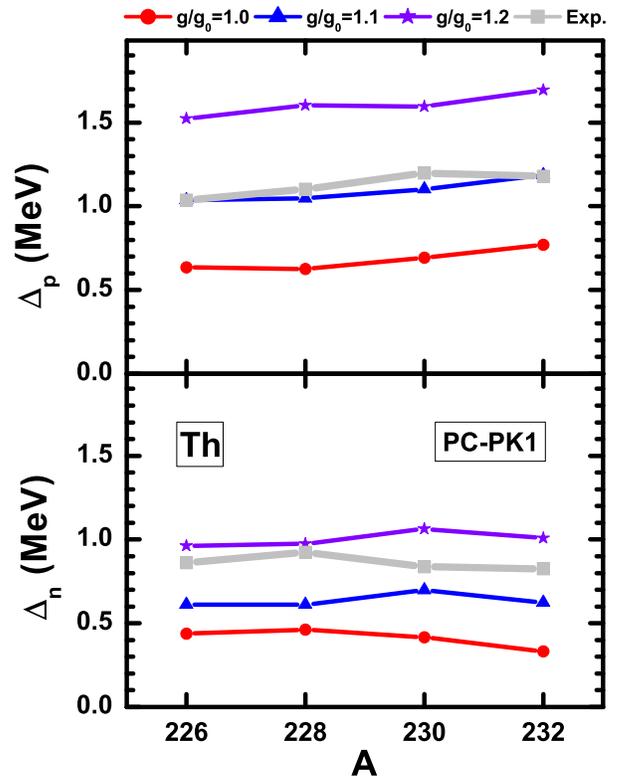}
\caption{(Color online)~\label{fig:Th_md}%
The odd-even differences of binding energies of Th isotopes computed with the MDC-RMF model, 
using the functional PC-PK1. Results obtained using different pairing strength parameters:
$g/g_0=1.0$, 1.1, and 1.2, where $g_0 =728$ MeV$\cdot$fm$^3$, 
are plotted as dots, triangles, and stars respectively. 
The squares denote the experimental values obtained from Ref.~\cite{Audi2012_ChinPhysC36-1287}.
The calculation has been performed in the ADHO basis with $N_{f}=20$ shells.
}
\end{figure}

\begin{figure}
 \includegraphics[width=0.45\textwidth]{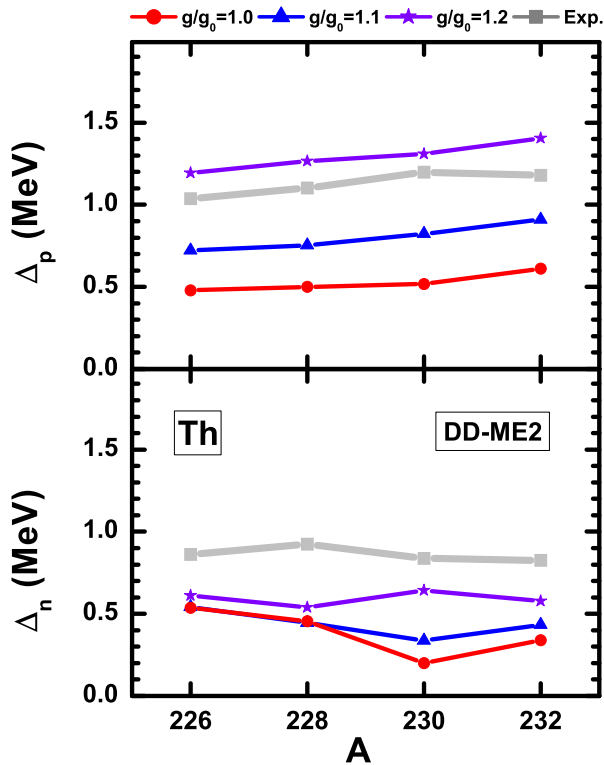}
\caption{(Color online)~\label{fig:Th_md_DDME2}%
The odd-even differences of binding energies of Th isotopes computed with the MDC-RMF model, 
using the functional DD-ME2. Results obtained using different pairing strength parameters:
$g/g_0=1.0$, 1.1, and 1.2, where $g_0 =728$ MeV$\cdot$fm$^3$, 
are plotted as dots, triangles, and stars respectively. 
The squares denote the experimental values obtained from Ref.~\cite{Audi2012_ChinPhysC36-1287}.
The calculation has been performed in the ADHO basis with $N_{f}=20$ shells.
}
\end{figure}

In the present work, since pairing correlations are treated in the BCS approximation, 
it is necessary to verify whether the strength of the pairing force is 
adequate for the mass region under consideration. 
We have therefore calculated the odd-even mass differences for the Th isotopes:
\begin{eqnarray}
% \begin{aligned}
  \Delta_{\rm n}(Z,N) &=& {1\over2} \left[ E(Z,N+1) + E(Z,N-1) - 2E(Z,N) \right], \nonumber \\
  \Delta_{\rm p}(Z,N) &=& {1\over2} \left[ E(Z+1,N) + E(Z-1,N) - 2E(Z,N) \right], \nonumber
% \end{aligned}
\end{eqnarray}
using different pairing strengths: $g/g_0=1.0$, 1.1 and 1.2, where $g_0 =728$ MeV$\cdot$fm$^3$. 
The results are shown in Figs.~\ref{fig:Th_md} and~\ref{fig:Th_md_DDME2}.
For the case of PC-PK1, when compared to experimental values obtained from 
Ref.~\cite{Audi2012_ChinPhysC36-1287}, one notices that the calculation 
with $g/g_0=1.0$ underestimates both proton and neutron odd-even mass differences considerably.
While the results obtained with $g/g_0=1.1$ nicely reproduce the empirical 
proton odd-even mass differences, the experimental neutron gaps are located 
between the values obtained using $g/g_0=1.1$ and $g/g_0=1.2$.
For the case of DD-ME2, the results obtained with both $g/g_0=1.0$ and $g/g_0=1.1$
underestimate proton and neutron odd-even mass differences.
The results obtained with $g/g_0=1.2$ reproduce the empirical proton odd-even mass differences 
quite well, and slightly underestimate the neutron odd-even mass differences. 
Since we do not wish to introduce additional model parameters by considering 
different pairing strengths for protons and neutrons, in the remainder of 
this study we will use $g/g_0=1.1$ both for protons and neutrons for PC-PK1 and
$g/g_0=1.2$ for DD-ME2.

Multidimensional self-consistent deformation-constrained calculations are very
time-consuming but, on the other hand, to accurately locate the saddle points
it is necessary to consider several shape degrees of freedom. 
Therefore, to locate the fission paths, in the present study we adopt the approach 
used in Refs.~\cite{Lu2012_PRC85-011301R, Lu2014_PRC89-014323}: 
(1) From the two-dimensional energy surfaces on the $(\beta_{20},\beta_{30})$ plane 
shown in Fig.~\ref{fig:pes}, and calculated with $N_{f} = 16$ and $g/g_0=1.0$, 
one approximately identifies the lowest fission path;
(2) In a second step, one-dimensional constrained calculations with $N_{f} = 20$ 
and $g/g_0=1.1$ for PC-PK1 and $g/g_0=1.2$ for DD-ME2 are performed along this 
approximate fission path for nuclei with apparent third minima. 
In these calculations additional nonaxial shapes are allowed around the second barrier.

\subsection{\label{subsec:1d_3rd}The third minima and barriers}

\begin{figure*}
 \includegraphics[width=0.8\textwidth]{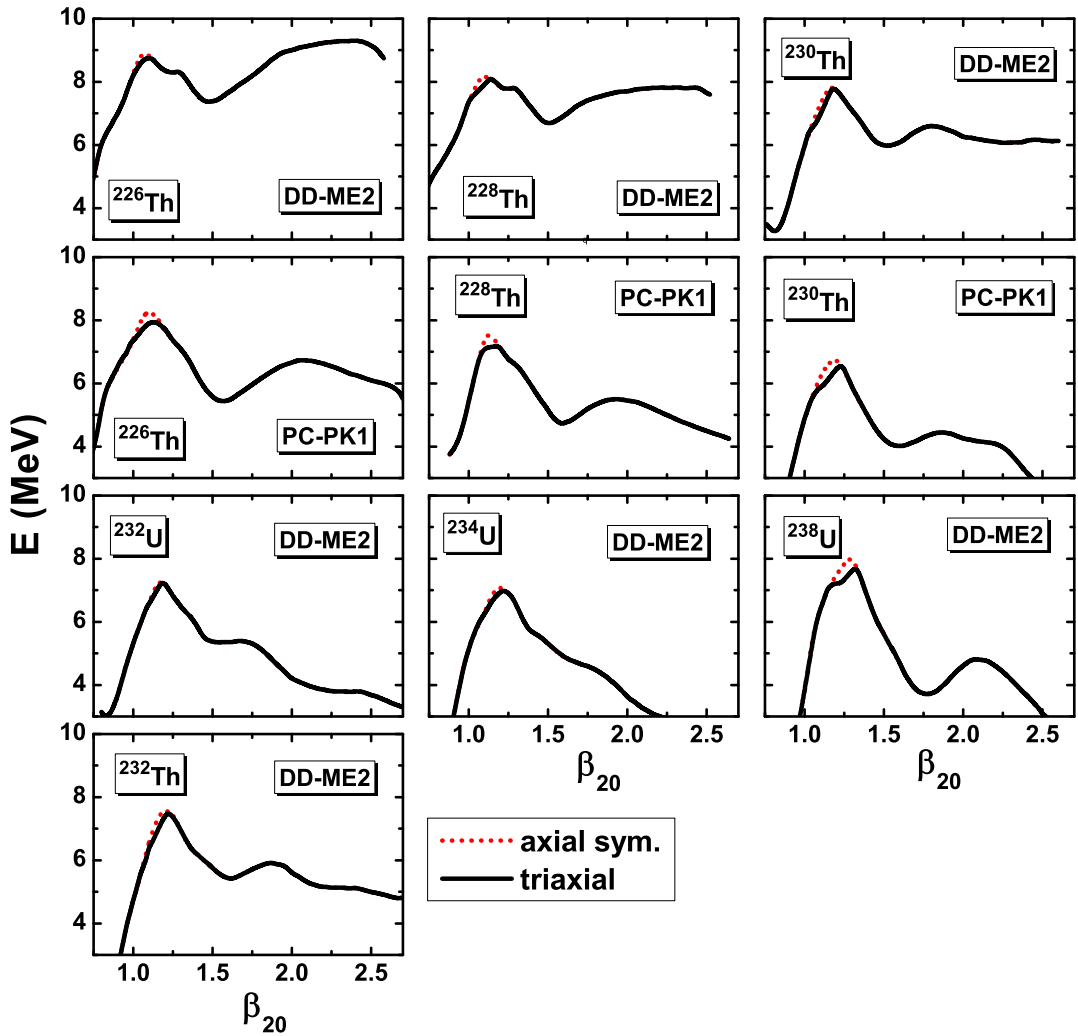}
\caption{(Color online)~\label{fig:Pathall}%
MDC-RMF energy curves of $^{226,228,230,232}$Th and $^{232,234,238}$U.
The calculation has been performed in the ADHO basis with $N_{f}=20$ shells 
and the pairing strength parameter $g/g_0=1.1$ for PC-PK1 and $g/g_0=1.2$ for DD-ME2.
For each nucleus the energy is normalized with respect to the binding energy
at the absolute minimum.}
\end{figure*}

By examining the two-dimensional PES's shown in Fig.~\ref{fig:pes}, 
one notices that third minima and barriers appear for $^{226,228,230,232}$Th 
and $^{232,234,238}$U. 
For these nuclei the energy curves along the lowest static fission path, 
calculated in the ADHO basis with $N_{f}=20$ shells and with the enhanced pairing 
strength $g/g_0=1.1$ for PC-PK1 and $g/g_0=1.2$ for DD-ME2, 
are shown in Fig.~\ref{fig:Pathall}.
In the vicinity of the second saddle point, MDC-RMF calculations are performed
with and without the inclusion of triaxial quadrupole shapes 
and both results are displayed for comparison.
As already reported in Ref.~\cite{Lu2012_PRC85-011301R}, the inclusion of 
triaxial configurations in addition to the axial octupole deformation, modifies 
the shape and height of the second fission barrier.
We should note that the effect of triaxiality on the second barrier
sensitively depends on the strength of the pairing interaction.
For instance, for the case $g/g_0=1.0$ (cf. Fig.~\ref{fig:Th226Path}), the inclusion of
triaxial configurations lowers the second fission barrier of $^{226}$Th by 0.72 MeV,
whereas for $g/g_0=1.1$ (cf. Fig.~\ref{fig:Pathall}) the second barrier
of $^{226}$Th is only lowered by 0.33 MeV. 
By further increasing the pairing strength the influence of nonaxial deformations 
on the binding energy near the second fission barrier may be further reduced 
\cite{Zhao2014_in-prep2}.
The energies of the third minima, the heights of the second and third barriers, 
and the depths of the third wells $\Delta E \equiv B_\mathrm{III} - E_\mathrm{III}$,
together with empirical parameters \cite{Capote2009_NDS110-3107,
Blons1988_NPA477-231,Yoneama1996_NPA604-263,Csige2009_PRC80-011301R,
Krasznahorkay1999_PLB461-15,Csige2013_PRC87-044321} are included
in Table~\ref{tab:parameter}.

\begin{table}
\caption{\label{tab:parameter} %
Excitation energies (in MeV) of the second saddle point $B_{\rm II}$, 
the third minimum $E_{\rm III}$, and the third saddle point $B_{\rm III}$,
with respect to the deformed mean-field equilibrium state for
$^{226,228,230,232}$Th and $^{232,234,238}$U, obtained from MDC-RMF calculations.
$\Delta E \equiv B_\mathrm{III} - E_\mathrm{III}$ 
denotes the depth of the third well relative to the third barrier.
The empirical values (denoted by ``Emp'')
are from Refs.~\cite{Capote2009_NDS110-3107,
Blons1988_NPA477-231,Yoneama1996_NPA604-263,Csige2009_PRC80-011301R,
Krasznahorkay1999_PLB461-15,Csige2013_PRC87-044321}.
}
%\footnotesize
\begin{ruledtabular}
\begin{tabular}{llllllcc}
%\begin{tabular*}{170mm}{@{\extracolsep{\fill}}ccccccccc}
 Nucleus     & Parameters & $B_{\rm II}$ & $E_{\rm III}$ & $B_{\rm III}$ & $\Delta E$ \\ \hline
 $^{226}$Th  & DD-ME2     & 8.76         & $7.37$        & $9.31$        & $1.94$     \\
             & PC-PK1     & 7.94         & $5.44$        & $6.73$        & $1.29$     \\
 $^{228}$Th  & DD-ME2     & 8.16         & $6.69$        & $7.82$        & $1.13$     \\
             & PC-PK1     & 7.19         & $4.72$        & $5.50$        & $0.78$     \\
 $^{230}$Th  & DD-ME2     & 7.84         & $5.97$        & $6.60$        & $0.63$     \\
             & PC-PK1     & 6.56         & $4.01$        & $4.45$        & $0.44$     \\
             & Emp~\cite{Capote2009_NDS110-3107} % P.3181, Table XI
                          & 6.80         &               &               &            \\
             & Emp~\cite{Blons1988_NPA477-231}   % Table 5
                          & 5.75         & $5.55$        & $6.45$        & $0.90$     \\
 $^{232}$Th  & DD-ME2     & 7.53         & $5.42$        & $5.92$        & $0.50$     \\
             & Emp~\cite{Capote2009_NDS110-3107} % P.3181, Table XI
                          & 6.70         &               &               &            \\
             & Emp~\cite{Yoneama1996_NPA604-263} % Table 1
                          & 6.50         & $5.40$        & $5.70$        & $0.30$      \\
 $^{232}$U   & DD-ME2     & 7.25         & \textemdash   & \textemdash   & \textemdash \\
             & Emp~\cite{Capote2009_NDS110-3107} % P.3181, Table XI
                          & 5.40         &               &               &             \\
             & Emp~\cite{Csige2009_PRC80-011301R}% Table I, the first row
                          & 4.91         & $3.20$        & $6.02$        & $2.82$      \\
 $^{234}$U   & DD-ME2     & 7.01         & \textemdash   & \textemdash   & \textemdash \\
             & Emp~\cite{Capote2009_NDS110-3107} % P.3181, Table XI
                          & 5.50         &               &               &            \\
             & Emp~\cite{Krasznahorkay1999_PLB461-15}
                          &              & $3.1 $        &               &            \\
 $^{238}$U   & DD-ME2     & 7.70         & $3.70$        & $4.81$        & $1.11$     \\
             & Emp~\cite{Capote2009_NDS110-3107} % P.3181, Table XI
                          & 5.50         &               &               &            \\
             & Emp~\cite{Csige2013_PRC87-044321} % Table II
                          & 5.6          & $3.6 $        & $5.6 $        & $2.0 $     \\
\end{tabular}
\end{ruledtabular}
\end{table}

For $^{226}$Th the functional DD-ME2 (PC-PK1) predicts the third minimum at
7.37 (5.44) MeV above the deformed ground state.
With DD-ME2 the third barrier is slightly higher than the second one,
whereas the opposite is obtained with PC-PK1.
The depth of the third well computed with DD-ME2 and PC-PK1 is 1.94 MeV
and 1.29 MeV, respectively. 
In the case of $^{228}$Th the third minimum is shallower:
1.13 MeV for DD-ME2 and 0.78 MeV for PC-PK1.
Among the nuclei considered here, these two isotopes exhibit the most pronounced third minima.
The fission barrier parameters for $^{230}$Th deduced in Ref.~\cite{Blons1988_NPA477-231} 
are $E_{\rm III}=5.55$ MeV
and $B_{\rm III}=6.45$ MeV.
MDC-RMF calculations with the functional DD-ME2 reproduce these values,
whereas they are underestimated by about 2 MeV by PC-PK1.
For $^{232}$Th the third minimum appears on the PES calculated with DD-ME2 but not
on the one obtained with PC-PK1.
The electron-induced fission cross section measurement for $^{232}$Th indicates 
a shallow third minimum with a depth of about 0.30 MeV \cite{Yoneama1996_NPA604-263}.
The theoretical result obtained with DD-ME2 (0.50 MeV) is consistent with this empirical value.
In general, Table~\ref{tab:parameter} shows that the depth of the third minimum 
in the Th isotopes decreases with increasing the neutron number.

For the U isotopes MDC-RMF calculations with the functional DD-ME2
predict the existence of the third minimum in $^{238}$U.
The depth of the third well is 1.11 MeV, which is smaller than the empirical value 
2.0 MeV~\cite{Csige2013_PRC87-044321}.
The reason for that is shown in Table~\ref{tab:parameter},
where one notices that the calculated energy of the third minimum (3.70 MeV)
is close to the empirical value (3.6 MeV),
but the theoretical height of the third barrier (4.81 MeV) is much smaller than  
the empirical value 5.6 MeV.
%For $^{232}$U the depth of the third well is 0.28 MeV, which is much smaller than
%the empirical value 2.82 MeV \cite{Csige2009_PRC80-011301R}. 
%The reason for that is shown in Table~\ref{tab:parameter},
%where one notices that the calculated height of the third barrier (5.89 MeV) 
%is close to the empirical value (6.02 MeV),
%but the theoretical energy of the third minimum (5.61 MeV) is considerably
%larger than the corresponding empirical excitation energy (3.20 MeV). 
%The calculated third well becomes even more shallow in $^{234}$U, 
%The calculated third well almost disappears in $^{234}$U, 
%whereas it is considerably deeper in $^{238}$U but not quite as deep as the
%empirical value.

\begin{figure*}
 \includegraphics[width=0.8\textwidth]{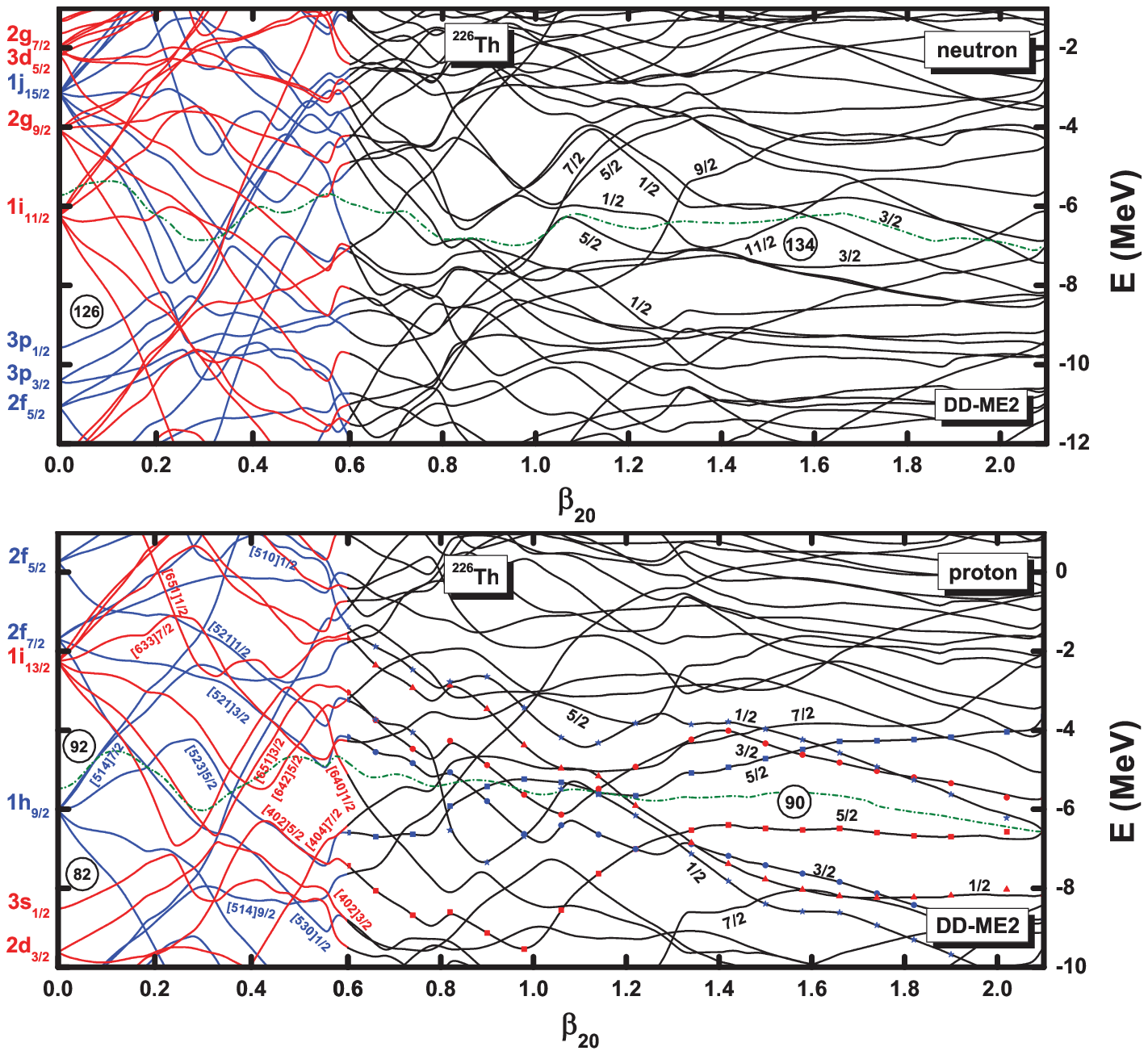}
\caption{(Color online)~\label{fig:lev}%
The neutron (upper panel) and proton (lower panel) single-particle levels
of $^{226}$Th near the Fermi surface along the static fission path.
For $\beta_{20} \leq 0.6$, only reflection-symmetric deformations are considered
and the red (blue) curves represent positive (negative) parity states.
When $\beta_{20} > 0.6$, the octupole deformation $\beta_{30}$ has a non-vanishing
value and parity is not a good quantum number.
The dash-dotted (green) curves denote the Fermi energy, and
the red (blue) symbols in the lower panel are used to guide the eye.
The MDC-RMF calculation has been performed with the functional DD-ME2.}
\end{figure*}

\subsection{Single-nucleon level structure}

The appearance of a hyperdeformed minimum must have its origin in the
single particle level structure.
Since $^{226}$Th displays the most pronounced third minimum among the nuclei
considered in this work, we will analyze the neutron and proton deformed
single-particle levels of this isotope.

In Fig.~\ref{fig:lev} we display the neutron and proton deformed single-particle 
levels of $^{226}$Th near the Fermi surface along the static fission path, 
as functions of the quadrupole deformation $\beta_{20}$. 
The levels are obtained in a MDC-RMF calculation using the functional DD-ME2.
The quadrupole deformation of the superdeformed minimum of $^{226}$Th is predicted
at $\beta_{20} \approx 0.6$.
For $\beta_{20} \leq 0.6$, octupole deformations are not considered,
and thus parity is conserved.
When $\beta_{20}>0.6$, the octupole deformation $\beta_{30}$ is nonzero and
the parity cannot be considered as a good quantum number for large deformations.
Furthermore, around the second saddle point triaxial deformations also play a role
and the third component of the angular momentum is not conserved either.
This results in a very complex single-particle level scheme around the second barrier.
Since in this study we are mainly interested in single-particle levels in the region
of the third minimum and the third barrier, in Fig.~\ref{fig:lev} we only plot results
obtained by imposing axial symmetry.

The hyperdeformed minimum of $^{226}$Th is located at 
$\beta_{20}\sim1.5$ and $\beta_{30}\sim0.7$.
By examining the neutron single-particle levels around $\beta_{20}=1.5$,
in the upper panel of Fig.~\ref{fig:lev} one notices a region of low
level density near the Fermi surface, even though the energy gap is not large.
For protons, as shown in the lower panel of Fig.~\ref{fig:lev},
a large energy gap is clearly visible at $Z=90$ in the region $\beta_{20} \approx 1.5$.
Therefore, the formation of the third minimum on the PES of $^{226}$Th is mainly caused
by the large proton gap at the Fermi surface.
Many single-particle states around the proton Fermi level are involved
in the formation of the energy gap at $Z=90$.
These states are dotted with red (blue) symbols in the lower panel of
Fig.~\ref{fig:lev}, and labeled with $\Omega$, i.e., the third component
of the angular momentum.

\section{\label{sec:summary}Summary}

We have analyzed the energy surfaces of light even-even Th and U isotopes
using the multidimensionally-constrained relativistic mean field (MDC-RMF) approach.
Calculations have been performed with two relativistic density functionals: 
PC-PK1 with point-coupling effective interactions that include higher order terms,
and the meson-exchange functional DD-ME2 with density-dependent meson-nucleon couplings.
Pairing correlations are taken into account in the BCS approximation with a separable
pairing force of finite range.

In a first step we have examined the two-dimensional PES's of $^{226,228,230,232}$Th 
and $^{232,234,236,238}$U, and located the third minima on the energy maps. 
By analyzing the resulting energy curves along the lowest static fission path,
the energies of the second barrier, the third minimum, and the
third barrier have been determined. 
In calculations with the functional DD-ME2, a third potential barrier is 
predicted in all Th nuclei and $^{238}$U.
The third well in $^{230,232}$Th is very shallow with a
depth of less than 1~MeV, whereas the third well in $^{226,228}$Th 
and $^{238}$U is rather deep. 
The functional PC-PK1 predicts a third barrier only in $^{226,228,230}$Th.
Therefore we note that the occurrence of a third barrier in constrained
mean-field calculations of PES's of actinides depends on the
specific choice for the energy density functional.

Insights into the origin of the third minimum on the PES have been obtained 
by examining the neutron and proton deformed single-particle levels
of $^{226}$Th near the Fermi surface along the static fission path, 
as functions of the quadrupole deformation $\beta_{20}$.
We have shown that the formation of the third minimum is facilitated by the
appearance of the $Z=90$ proton energy gap in the region
$\beta_{20} \approx 1.5$ and $\beta_{30} \approx 0.7$
and this gap originates from several pairs of single-proton states
in the vicinity of the Fermi surface.

\acknowledgements
We are grateful to the Kavli Institute for Theoretical Physics China (KITPC) at
the Chinese Academy of Sciences where this work was initiated.
We thank Roberto Capote, M. Kowal, U.-G. Mei{\ss}ner, P. Ring, J. Skalski, Kai Wen,
and Zhen-Hua Zhang for helpful discussions.
This work has been supported by
the National Key Basic Research Program of China (Grant No. 2013CB834400),
the National Natural Science Foundation of China (Grants
No. 11121403,
No. 11175252,
No. 11120101005,
No. 11211120152, and
No. 11275248),
and
the Knowledge Innovation Project of the Chinese Academy of Sciences (Grant No. KJCX2-EW-N01).
The computational results presented in this work have been obtained on
the High-performance Computing Cluster of SKLTP/ITP-CAS and
the ScGrid of the Supercomputing Center, Computer Network Information Center of
the Chinese Academy of Sciences.

%\bibliographystyle{apsrev4-1}
%\bibliography{/home/bnlu/Desktop/Works/Documents/Personal.Files/Tetrahedra}
%\bibliography{/home/phantom/My_DATA/MyWork/Material/Paper/Nuclear-Phys/nuclear,/home/phantom/My_DATA/MyWork/Material/Paper/Nuclear-Fission/Nuclear-Fission}
%\bibliography{../../../information/refs/JabRef/sgzhou}


\end{document}